\documentclass[twocolumn,aps,showpacs]{revtex4-1}
\usepackage{amsfonts}
\usepackage{amsmath}
\usepackage{graphicx}
\usepackage{bm}
\usepackage{amssymb}
\usepackage{dcolumn}
\usepackage{color}

\begin{document}

\title{Quantum Renormalization of the Spin Hall Effect}

\author{Bo Gu$^{1,2}$, Jing-Yu Gan$^{3}$, Nejat Bulut$^{4}$,
Timothy Ziman$^{5}$, Guang-Yu Guo$^{6,7}$,
Naoto Nagaosa$^{8,9}$, and Sadamichi Maekawa$^{1,2}$}

\affiliation{$^1$Advanced Science Research Center, Japan Atomic Energy Agency, Tokai 319-1195, Japan
\\ $^{2}$JST, CREST, 3-Sanbancho,Chiyoda-ku, Tokyo 102-0075, Japan
\\ $^{3}$Institute of Physics, Chinese Academy of Sciences, P.O. Box 603, Beijing 100190, China
\\ $^{4}$Department of Physics, Izmir Institute of Technology, Urla 35430, Turkey  
\\ $^{5}$CNRS and Institut Laue Langevin, Bo\^\i te Postale 156, F-38042 Grenoble Cedex 9, France
\\ $^{6}$Graduate Institute of Applied Physics, National Chengchi University, Taipei 116, Taiwan
\\ $^{7}$Department of Physics, National Taiwan University, Taipei 106, Taiwan
\\ $^{8}$Department of Applied Physics, The University of Tokyo, Tokyo 113-8656, Japan
\\ $^{9}$Cross-Correlated Materials Research Group and 
Correlated Electron Research Group, RIKEN-ASI, Wako 315-0198, Japan}

\begin{abstract}
By quantum Monte Carlo simulation of a realistic multiorbital
Anderson impurity model, we study the
spin-orbit interaction (SOI) of an Fe impurity in Au host metal. 
We show, for the first time, that the SOI is strongly renormalized by the 
quantum spin fluctuation. Based on this mechanism, we can explain why the gigantic
spin Hall effect in Au with Fe impurities was observed in recent experiment, 
while it is not visible in the anomalous Hall effect. 
In addition, we show that the SOI is strongly renormalized by the
Coulomb correlation U. Based on this picture, we can explain past
discrepancies in the calculated orbital angular momenta for an Fe impurity 
in an Au host. 
\end{abstract}

 \maketitle

When a magnetic impurity with $d$ orbitals is put into a normal metal
with $s$- or $p$-conduction bands, it forms a virtual bound state
hybridized with the conduction states, leading to a broadening
with width $\Delta$ (hybridization energy) which is typically of
the order of eV. This energy scale competes with the
Coulomb energy $U$ between the electrons in the $d$ orbitals,
which is also of the order of eV.
The latter tends to produce the spin moment, while the
former induces quantum fluctuation of that
spin moment, leading to the spin singlet.
The competition between these two interactions
defines the time scale, or equivalently the energy scale,
of the quantum spin fluctuation, i.e., the Kondo
temperature $T_K$, which can be much lower than
both $\Delta$ and $U$ 
\cite{AuFe-R}.
In the language of the
renormalization group, this is described
by the energy dependent scaling of the
various operators. This means that
even a weak interaction can be amplified
in the low energy or temperature scale and
compete with the much larger energy scale
due to the correlation $U$. 

In real systems, the $d$ orbitals have fivefold
degeneracy and these orbital degrees of freedom
have often been neglected in the analysis
of the experimental results. The reasoning is
that either the crystal field splitting $\Delta \varepsilon$
is  much smaller than the hybridization energy $\Delta$,
or that it is in the limit with $\Delta \varepsilon$  much
larger than the Kondo temperature $T_K$.
However, the naive comparison between the
bare interaction strengths is dangerous since these
are scale-dependent running coupling constants.
Similar nontrivial behavior can also be expected 
for the relativistic spin-orbit interaction.

Orbital degrees of freedom in the impurity scattering
lead to intriguing phenomena such as the anomalous Hall
effect (AHE) and the spin Hall effect (SHE).
A charge current perpendicular to the applied electric field
is produced in ferromagnetic metals (AHE), while a spin current
rather than a  charge current, is induced in semiconductors
and metals without magnetism (SHE).
An extrinsic mechanism of these two effects arises
from the skew scattering, i.e., spin-dependent deflection
of the scattered electrons due to the spin-orbit interaction.
It has been shown in the case of the AHE that the resonant
skew scattering due to the virtual bound $d$ states leads
to a large Hall angle, of the order of 0.01 
\cite{AuFe-AHE}, compared with
the typical value of $10^{-3}$. In the AHE, the
spin fluctuation is quenched, owing to the ferromagnetism
or the external magnetic field, and the ratio of the
spin-orbit interaction $\lambda$ and $\Delta$ basically
determines the Hall angle. On the other hand,
in the SHE, the spin fluctuation is active and the
Kondo physics can be relevant to the resonant
skew scattering. Therefore, a crucial question
is whether the Kondo effect and quantum spin fluctuation
can produce an even larger spin Hall angle compared with the AHE.

An important clue to this question comes from an experiment by Seki
et al. \cite{AuFe-SHE} on Au/FePt, in which a spin Hall
angle of 0.114 was observed. Motivated by this experiment, some of the present
authors studied the Fe impurity in Au by a first-principles
calculation \cite{AuFe-GMN}. The Kondo effect of Fe impurities is a
historic problem, with a low $T_K$ around 0.4K
\cite{AuFe-Kondo}, high electric resistance at room temperature
\cite{AuFe-R}, and an AHE with Hall angle of the
order of 0.01 \cite{AuFe-AHE}. However,  a simple  fivefold degeneracy of the
orbitals has been  assumed to analyze the experiments. In Ref.
\cite{AuFe-GMN}, on the other hand, the orbital-dependent Kondo
effect of Fe in Au was proposed to explain the nature of the
experimentally observed giant spin Hall signals \cite{AuFe-GMN}: It
was argued that the $e_g$ orbitals of Fe are in the Kondo
limit and $t_{2g}$ orbitals are in the mixed-valence region. The
enhancement  of the  spin-orbit interaction by  electron correlation in the
$t_{2g}$ orbitals leads to the giant spin Hall effect. However this
proposal has been challenged theoretically by
Ref.\cite{AuFe-Costi}, which suggests an effective 3-channel Kondo
model, involving local and band electrons of $t_{2g}$ symmetry, and
also experimentally by the x-ray MCD \cite{AuFe-Brewer}, which
obtained a rather small value of the orbital angular momentum in
contrast to the large value calculated in Ref.\cite{AuFe-GMN}.

In order to resolve this confusing situation and quantify the mechanism
for the enhanced spin Hall effect, it is essential to
treat the quantum fluctuations of the spins and orbitals
systematically. This is impossible in the first-principles
calculation, which  assumes ordered spin and orbital moments.
In this Letter, we overcome this difficulty by using
the Hirsch-Fye quantum Monte Carlo (QMC) simulation \cite{QMC}, 
combined with the
density functional theory (DFT)\cite{DFT1,DFT2}, to study the renormalization
due to correlation effects.
First, a single-impurity multiorbital Anderson model
\cite{Anderson} is formulated within the DFT 
for determining the host band structure, the
impurity levels, and the impurity-host hybridization. Second,
the magnetic behaviors of the Anderson impurity at finite
temperatures are calculated by QMC. 

The single-impurity multiorbital Anderson model is defined as
\begin{eqnarray}
  H&=&\sum_{\textbf{k},\alpha,\sigma}\epsilon_{\alpha}(\textbf{k})
  c^{\dag}_{\textbf{k}\alpha\sigma}c_{\textbf{k}\alpha\sigma}
   +\sum_{\textbf{k},\alpha,\xi,\sigma}(V_{\xi\textbf{k}\alpha }
    d^{\dag}_{\xi\sigma} c_{\textbf{k}\alpha\sigma} + H.c.) \notag\\
  &+& \sum_{\xi,\sigma}\epsilon_{\xi}n_{\xi\sigma}
  +U\sum_{\xi}n_{\xi\uparrow}n_{\xi\downarrow} \notag\\
   &+& \frac{U^{\prime}}{2}\sum_{\xi\neq\xi',\sigma,\sigma^{\prime}}
     n_{\xi\sigma}n_{\xi'\sigma^{\prime}} 
   - \frac{J}{2}\sum_{\xi\neq\xi',\sigma}n_{\xi\sigma}n_{\xi'\sigma},
\label{E-Ham}
\end{eqnarray}
where $c^{\dag}_{\textbf{k}\alpha\sigma}$
($c_{\textbf{k}\alpha\sigma}$) is the creation (annihilation)
operator of the conduction electron with wavevector $\textbf{k}$ and spin
$\sigma$ in the band $\alpha$, $d^{\dag}_{\xi\sigma}$
($d_{\xi\sigma}$) is the creation (annihilation) operator of the
localized electron at the impurity site with orbital $\xi$ and spin $\sigma$,
and $n_{\xi\sigma}=d^{\dag}_{\xi\sigma}d_{\xi\sigma}$.
The host energy band $\epsilon_{\alpha}(\textbf{k})$, the impurity
energy levels $\epsilon_{\xi}$, and the impurity-host
hybridization $V_{\xi\textbf{k}\alpha}$, as the one-body problems,
can be properly obtained by the DFT calculations.
$U$ ($U'$) is the on-site Coulomb repulsion within (between) the orbitals of the
impurity, and $J$ is the Hund coupling between the orbitals of the impurity.
These many-body interactions can be exactly treated by the QMC calculations,
and thus it becomes possible to accurately study the quantum fluctuations of the spins
and orbitals of the impurity.
Considering the parameters used in the previous
calculations for Fe in Au \cite{AuFe-GMN}, and the
relationship $U$ = $U^{\prime}$ + 2$J$ \cite{SM-TT}, in our following QMC calculations,
we use the values of $U$ = 5 eV, $J$ = 0.9 eV, and $U^{\prime}$ = 3.2 eV for most cases,
but  we shall vary the  values for a few cases in order to clarify the role of
correlations and reconcile with past calculations.

Our DFT calculations are done by the 
code {\sc Quantum-ESPRESSO} \cite{ESPRESSO}.
To calculate the impurity-host hybridization,
we consider the supercell Au$_{26}$Fe, where
the exchange-correlation interactions are described by the
Perdew-Zunger local density approximation (LDA), and the
electron-ion interactions are represented by the
Rabe-Rappe-Kaxiras-Joannopoulos ultrasoft pseudopotentials.
4$\times$4$\times$4 Monshorst-Pack $\textbf{k}$ points are used. 
Kinetic energy cutoff for wavefunctions and charge density
are taken as 30 and 300 Ry, respectively.

\begin{figure}[tbp]
\includegraphics[width = 8.5 cm]{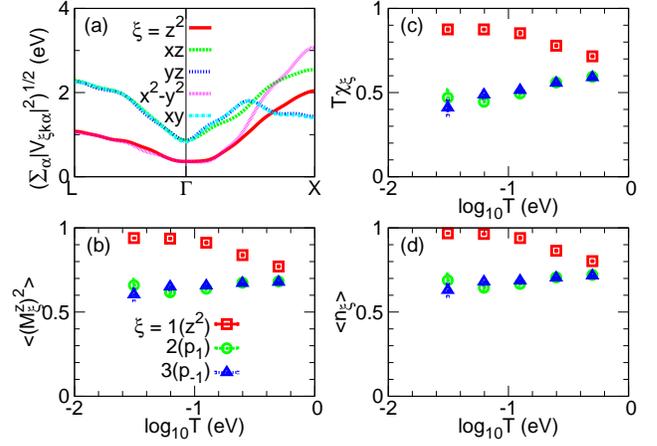}
\caption{ (color online) (a) Hybridization between the $\xi$ orbitals 
of an Fe impurity and an Au host. We show the LDA result including five $d$ orbitals.
(b) Temperature dependence of the square of the magnetic moment
$\langle(M^z_{\xi})^2\rangle$, (c) the temperature times susceptibility
$T\chi_{\xi}$, and (d) the occupation number $\langle n_{\xi}\rangle$
of the $\xi$ orbitals of the Fe impurity.
(b)-(d) are the QMC results including three $d$ orbitals. See text for details.}
\label{F-3orb}
\end{figure}

Figure \ref{F-3orb}(a) shows the hybridization between $\xi$ orbitals 
of Fe impurity and Au host, where
the hybridization matrix element has the form of
$V_{\xi\textbf{k}\alpha}\equiv\langle\varphi_{\xi}
|H_0|\Psi_{\alpha}(\textbf{k})\rangle$ =
$ \frac{1}{\sqrt{N}}
\sum_{p,\textbf{n}}e^{i \textbf{k}\cdot\textbf{n}}a_{\alpha
p}(\textbf{k})
\langle\varphi_{\xi}|H_0|\varphi_{p}(\textbf{n})\rangle$.
$H_0$ is the one-particle part of Eq. (\ref{E-Ham}), $\varphi_{\xi}$ is the
$\xi$ state of Fe impurity, and $\Psi_{\alpha}(\textbf{k})$ is the
Au host state with wavevector $\textbf{k}$ and band index $\alpha$,
which is expanded by atomic orbitals $\varphi_{p}(\textbf{n})$ with
orbital index $p$ and site index $\textbf{n}$, $a_{\alpha
p}(\textbf{k})$ are expansion coefficients,
$\langle\varphi_{\xi}|H_0|\varphi_{p}(\textbf{n})\rangle$ are mixing integrals,
and $N$ is the number of lattice sites.
It is observed that, at the $\Gamma$ point ($\textbf{k}$=0), the
hybridization value of the $\xi$
= $e_g$($z^2$,$x^2$-$y^2$) orbitals of the Fe impurity is smaller than
that of the $\xi$ = $t_{2g}$ ($xz$,$yz$,$xy$) orbitals.

Based on the above DFT/LDA calculation, 
we can determine approximate impurity levels $\epsilon_{\xi}$= -1.9 eV for
$\xi$ = $e_g$ ($z^2$, $x^2$-$y^2$) and $\epsilon_{\xi}$= -1.8
eV for $\xi$ = $t_{2g}$ ($xz$,$yz$,$xy$) with zero Fermi energy,
where the Coulomb terms included in the LDA calculations have been subtracted 
as in Refs.~\cite{Shift1, Shift2}.
The crystal field splitting $\Delta \varepsilon$ = 0.1 eV is
in agreement with
the previous LDA calculations \cite{AuFe-GMN,AuFe-Costi}.

To study the renormalization due to the correlation effect
for (i) the crystal field splitting
and (ii) the relativistic spin-orbit interaction,
the magnetic behaviors of Eq.~(\ref{E-Ham}) are calculated by
the Hirsch-Fye QMC simulation for the infinite host system \cite{QMC}.
Owing to calculation constraints,
we simplify it to a three-orbital model with one
$e_g$ and two $t_{2g}$ orbitals, which captures the essential physics.
The $e_g$ orbital is arbitrarily chosen as $z^2$,
and the two $t_{2g}$ orbitals are chosen
as $p_1$ and $p_{-1}$, where the notation corresponds to the
transformational properties of $t_{2g}$
orbitals equivalent to effective $p$ orbitals \cite{T2gP}: $p_1\equiv
-\frac{1}{\sqrt{2}}(xz-iyz)$, $p_0\equiv -ixy$, and $p_{-1}\equiv
-\frac{1}{\sqrt{2}}(xz+iyz)$.
The following results are obtained with more
than 10$^{5}$ Monte Carlo sweeps and the Matsubara time step
$\Delta\tau=0.25$. 

Figures \ref{F-3orb}(b)-(d) show the temperature dependence of the
square of the magnetic moment $\langle(M^z_{\xi})^2\rangle$, the
temperature times susceptibility $T\chi_{\xi}$, and the
occupation number $\langle n_{\xi}\rangle$, which are defined as
$M^z_{\xi}$ = $ n_{\xi\uparrow} - n_{\xi\downarrow}$,
$\chi_{\xi}$ = $ \int_{0}^{\beta}d\tau \langle
M^z_{\xi}(\tau)M^z_{\xi}(0)\rangle$,
$n_{\xi}$ = $n_{\xi\uparrow} + n_{\xi\downarrow}$,
respectively, and the impurity levels $\epsilon_{\xi}$= -1.9
($\xi$=$z^2$) and -1.8 eV ($\xi$=$p_1$, $p_{-1}$) are used.
It is found that the $e_g$($t_{2g}$) orbital has
a larger (smaller) magnetic moment, 
a much larger (smaller) susceptibility with a much
smaller (larger) Kondo temperature, 
and a larger (smaller) occupation number. 
In addition, if we repeat the QMC calculation 
with the degenerate impurity levels by hand,
$\epsilon_{\xi}$= -1.85 eV for $\xi$=$z^2$, $p_1$ and $p_{-1}$,
nearly the same behaviors are observed.
Thus it is clear that the orbital-dependent Kondo effect
comes mainly from the renormalization of the impurity-host hybridization
by correlations.

\begin{figure}[tbp]
\includegraphics[width = 8.5 cm]{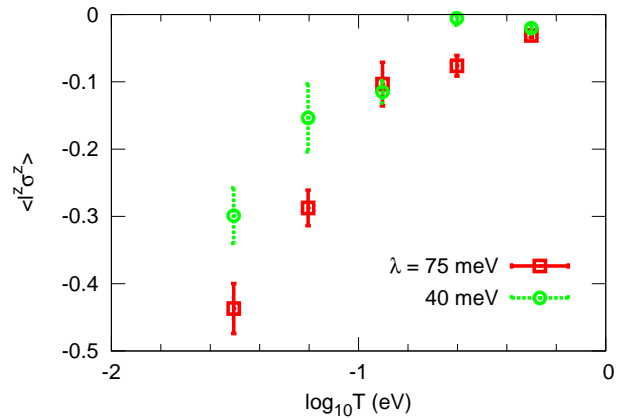}
\caption{(color online) Temperature dependence of spin-orbit interactions
for an Fe impurity in an Au host. The parameter $\lambda$ in Eq. (\ref{E-SOI})
is taken as a realistic 75 meV  \cite{AuFe-SOI}, and as 40 meV for comparison.
} \label{F-3so}
\end{figure}

Next, we study problem (ii), the renormalization of the
relativistic spin-orbit interaction due to the correlation.
For simplicity,  we consider only the $z$ component,
\begin{eqnarray}
H_{so}=(\lambda/2)\left(\ell^z\sigma^z\right),
\ell^z\sigma^z&\equiv& n_{2\uparrow}-n_{2\downarrow} -n_{3\uparrow}
+n_{3\downarrow},\label{E-SOI}
\end{eqnarray}
where $\textbf{l}^z$ is the 3$\times$3 $z$ component of the angular
moment matrix of $\ell=1$, and $\sigma^z$ is the 2$\times$2 $z$ component
of Pauli matrix. $\xi$ = 2(3) denotes the $p_1$($p_{-1}$) orbital.
We add Eq.(\ref{E-SOI}) to Eq.(\ref{E-Ham}),
where the parameters in Eq.(\ref{E-Ham}) are taken as the same values of
Figs. \ref{F-3orb}(b)-(d). For an Fe atom, the realistic value of the  spin-orbit
interaction  is  $\lambda$ = 75 meV
\cite{AuFe-SOI}. We also show the results with the smaller
value of $\lambda$ = 40 meV for comparison. Considering that the value of
$\lambda$ is 2 orders of magnitude less than that of impurity
energies $\epsilon_{\xi}$, it is not surprising that the temperature
dependence of the square of the magnetic moment
$\langle(M^z_{\xi})^2\rangle$, the  susceptibility
$T\chi_{\xi}$, and the occupation number $\langle n_{\xi}\rangle$
are nearly the same as those of Figs. \ref{F-3orb}(b)-(d). In contrast,
as displayed in Fig. \ref{F-3so}, a nonzero spin-orbit
correlation function $\langle \ell^z\sigma^z\rangle$ appears when
the spin-orbit interaction within the $t_{2g}$ orbitals of the Fe impurity
is included. At  temperature  $T$= 360 K, the lower limit of our
present calculations, we have $\langle \ell^z\sigma^z\rangle$$\cong$
-0.44 with $\lambda$ = 75 meV, and
$\langle \ell^z\sigma^z\rangle$$\cong$ -0.3 with $\lambda$ = 40 meV.

To compare the QMC results with the experiment,
we calculate the spin Hall angle $\gamma_s$ as follows.
Since we consider only two $t_{2g}$ orbitals with
the $z$-component of the orbital angular moment $\ell^z$ = $\pm$1,
the spin-orbit interaction within the $t_{2g}$ orbitals gives rise to
the difference in the occupation numbers between
the parallel ($n_P$) and antiparallel ($n_{AP})$ states of the spin and
angular momenta. These occupation numbers are related to the
phase shifts $\delta_P$ and $\delta_{AP}$,
respectively as $n_{P(AP)}$ = $\delta_{P(AP)}$$/\pi$.
These quantities can be estimated as
\begin{eqnarray}
\pi \langle\ell^z \sigma^z\rangle = \delta_P - \delta_{AP},
\pi \langle n_2\rangle + \pi \langle n_3\rangle = \delta_P + \delta_{AP},
\end{eqnarray}
which are given in Fig. 2 and 1(d), respectively.
Putting $\langle\ell^z \sigma^z\rangle$= -0.44 for $\lambda$= 75 meV,
and $\langle n_2\rangle$= $\langle n_3\rangle$= 0.65, we obtain
$\delta_P$= 1.35 and $\delta_{AP}$= 2.73.
Taking into account the estimate $\delta_1 \cong$ 0.1 of
the phase shift for p-wave scattering,
and applying the equation \cite{AuFe-GMN}
$\gamma_s$ = $6Im[(e^{-2i\delta_1}-1)(e^{2i\delta_{P}}-e^{2i\delta_{AP}})]$
$/[25-15\cos2\delta_P-10\cos2\delta_{AP}]$,
the spin Hall angle is thus obtained as $\gamma_s \cong$ 0.055,
comparable to that observed in recent experiment.

To further understand how correlations renormalize
the effective relativistic
spin-orbit interaction, we will now vary
the correlation energy and the spin polarization,
while keeping the temperature $T$ = 360 K and
spin-orbit interaction $\lambda$ = 75 meV fixed.

\begin{figure}[tbp]
\includegraphics[width = 8.5 cm]{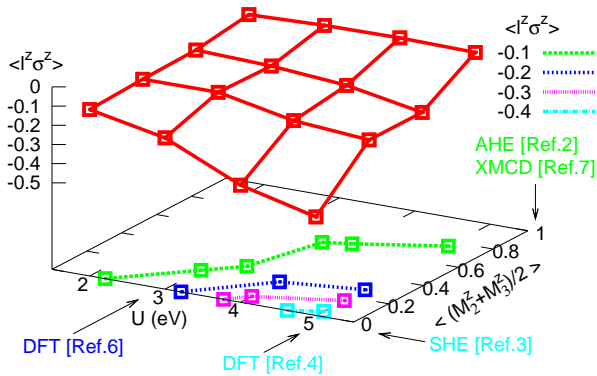}
\caption{(color online) For an Fe impurity in an Au host,
we show the spin-orbit interaction
$\langle \ell^z\sigma^z\rangle$ as a function
of correlation energy $U$ and spin polarization
$\langle (M^{z}_{2}+M^z_3)/2\rangle$. 
The QMC results of 
$\langle \ell^z\sigma^z\rangle$ are shown as red squares.
In the phase diagram plane
of $U$ to $\langle (M^{z}_{2}+M^z_3)/2\rangle$,
the different colors denote the different values of
$\langle \ell^z\sigma^z\rangle$.
Here, $T$ = 360 K and $\lambda$ = 75 meV are fixed.}
\label{F-phase}
\end{figure}

\emph{Effect of correlation U}:
For each $U$,
we keep $J/U$ = 0.9/5 and $U$ = $U^{'}$ +2$J$.
In Fig. \ref{F-phase}
with $\langle (M^{z}_{2}+M^z_3)/2\rangle$ = 0,
we see that the spin-orbit correlation function
$|\langle \ell^z\sigma^z\rangle|$ dramatically decreases
as $U$ decreases. This demonstrates clearly how
the correlation $U$  renormalizes the relativistic spin-orbit
interaction. The results also explain past discrepancies
in the calculated orbital angular momenta for an Fe
impurity in an Au host: A large value is calculated with $U$ = 5 eV
\cite{AuFe-GMN}, while much smaller  ones are obtained with $U$ = 3 or
2 eV \cite{AuFe-Costi,AuFe-Chadov}.
The virtual bound states of $t_{2g}$ have 
majority and minority parts, whose contributions to 
$\langle \ell^z\sigma^z\rangle$ are opposite in sign. 
Increasing $U$ will push up the minority part away from the Fermi energy, 
decrease its contribution to $\langle \ell^z\sigma^z\rangle$,
and hence increase the total
$\langle \ell^z\sigma^z\rangle$.

\emph{Effect of spin polarization}: To see clearly the effects of  quantum
fluctuations, it is instructive to study the effects of polarizing
the Fe impurity spin, which can quench quantum fluctuations of the spins and orbitals.
We apply a magnetic field $h$ by hand on the orbitals 1($z^2$),
2($p_1$), and 3($p_{-1}$) of the Fe impurity,
\begin{eqnarray}
H_{f}=-h\left(n_{1\uparrow}-n_{1\downarrow}
+n_{2\uparrow}-n_{2\downarrow}+n_{3\uparrow}-n_{3\downarrow}\right),
\label{E-t2geg}
\end{eqnarray}
where $h$ may be of a fraction of eV in order to
polarize the spins at the temperature of 360 K.
We add Eq. (\ref{E-t2geg}) to the previous Hamiltonian
with the parameters in Eq.(\ref{E-Ham})  taken
to be  the same as in Fig. \ref{F-3so}.
It is found that, in Fig. \ref{F-phase} with $U$ = 5 eV,
when $h$ is set up to 0.8 eV to make the spins
$\langle M^z_{\xi}\rangle$ of $\xi$ = 1($z^2$), 2($p_{1}$) and
3($p_{-1}$) gradually polarize near the saturated value 1,
the spin-orbit correlation function $|\langle \ell^z\sigma^z\rangle|$
gradually decreases to 0. Thus, the quantum fluctuation of the spin is crucially
important in  producing an even larger
spin Hall angle than the Hall angle of the AHE.
What happens is that the quantum fluctuation of the spin 
leads to the motional narrowing of the spin-orbit coupled 
virtual bound states, reducing the hybridization energy with the conduction
electrons and hence the energy width $\Delta$. 
So, the Hall angle will 
increase with decreasing $\Delta$ \cite{AuFe-AHE}. 

\emph{Phase diagram}: Combining  the above two parameters, our QMC results
for the spin-orbit correlation function $\langle \ell^z\sigma^z\rangle$
are noted as red squares in Fig. \ref{F-phase}.
In the phase diagram plane
of the correlation $U$ to the spin polarization $\langle
(M^{z}_{2}+M^z_3)/2\rangle$,
as also noted in Fig. \ref{F-phase},
the different colors note the different values of
$\langle \ell^z\sigma^z\rangle$.
It is seen that the areas with large spin-orbit correlation
should possess strong correlation energy $U$ and weak spin polarization.
The measurement of the orbital angular momentum by x-ray MCD involves
the application of a sum rule for  atomic wavefunctions: In order to
eliminate interference between different impurities, it is necessary
to polarize  with an external magnetic field at low temperatures\cite{AuFe-Brewer}.
The spin Hall angle measurement  \cite{AuFe-SHE} was done
in a spin-unpolarized state and at room temperature. From our phase diagram in Fig.
\ref{F-phase}, it is suggested that the orbital angular momentum in a spin
polarized state could be much smaller than that in 
a spin-unpolarized state. We emphasize that this phase diagram is obtained
for temperatures comparable to room temperature, as is relevant
for spin Hall applications; for quantitative comparison
of very low temperature measurements, such as x-ray MCD and phase decoherence,
the scales can be renormalized. 

To conclude, we have studied the
SHE for an Fe impurity in Au host by QMC simulation of a multiorbital 
Anderson impurity model.
Our phase diagram with respect to correlation and spin polarization shows that
large spin-orbit correlations occur for strong correlation $U$ and weak spin polarization,
which, in principle, may resolve the current, somewhat confusing, situation of experiments 
and calculations. More generally, we can, for the first time, quantify an essential difference
between the AHE and the SHE.
While we have focused on the specific case of Fe in Au, the results
suggest a more general mechanism, and constraints on parameters, for obtaining a large SHE by impurities.

This work was supported by 
Grant-in-Aids for Scientific Research in the priority area 
"Spin current" with No. 19048015, No. 19048009, No. 19048008, and No. 21244053,  
the Next Generation Super Computing Project,
the Nanoscience program from MEXT of Japan, 
the Funding Program for World-Leading Innovative R$\&$D on Science and Technology (FIRST Program),
and NSC of Taiwan.
T.Z. is grateful to the ICC-IMR.

\end{document}